\documentstyle[amssymb,aps,preprint]{revtex}

\def\be{\begin{equation}}
\def\ee{\end{equation}}
\def\bea{\begin{eqnarray}}
\def\eea{\end{eqnarray}}

\begin{document}
\title{Phase diagram of the {\it XXZ} chain with next-nearest-neighbor interactions}
\author{R. D. Somma and A. A. Aligia}
\address{Comisi\'{o}n Nacional de Energ{\'{\i }}a At\'{o}mica,\\
Centro At\'{o}mico Bariloche and Instituto Balseiro, 8400 S.C. de Bariloche,%
\\
Argentina}
\maketitle

\begin{abstract}
We calculate the quantum phase diagram of the {\it XXZ} chain with
nearest-neighbor (NN) $J_{1}$ and next-NN exchange $J_{2}$ with anisotropies 
$\Delta _{1}$ and $\Delta _{2}$ respectively. In particular we consider the
case $\Delta _{1}=-\Delta _{2}$ to interpolate between the {\it XX} chain ($%
\Delta _{i}=0$) and the isotropic model with ferromagnetic $J_{2}$. For $%
\Delta _{1}<-1$, a ferromagnetic and two antiferromagnetic phases exist. For 
$\left| \Delta _{i}\right| <1$, the boundary between the dimer and spin
fluid phases is determined by the method of crossing of excitation spectra.
For large $J_{2}/J_{1}$, this method seems to indicate the existence of a
second spin fluid critical phase. However, an analysis of the spin stiffness
and magnetic susceptibility for $\Delta _{1}=\Delta _{2}=1$ suggest that a
small gap is present.
\end{abstract}


\section{Introduction}

In recent years, several interesting quasi one dimensional magnetic systems
have been studied experimentally \cite{has,mot,col,miz}. Among them some
compounds containing CuO chains with edge sharing CuO$_{4}$ plaquettes are
expected to be described by the {\it XXZ} model with an important next-NN
exchange $J_{2}$ in comparison with the NN exchange $J_{1}$, and the latter
can also be negative if the Cu-O-Cu angle is near 90${{}^{\circ }}$ \cite
{miz}. In addition, there has been a renewed interest in this {\it XXZ}
model (which is equivalent to a zig-zag ladder) recently \cite
{bur,whi,all,ner,inco,hik,ger,hir1,meta,cab,hir2,itoi}. In particular, for
the {\it XX} chain ($\Delta _{1}=\Delta _{2}=0$), field theoretical methods
predicted a critical (gapless) phase with incommensurate correlations for $%
J_{2}>>J_{1}>0$. This seems confirmed by recent numerical work \cite{hik}.
The effects of magnetic field and magnetization jumps (metamagnetism) have
also been studied recently \cite{ger,hir1,meta,cab}. Metamagnetism is more
likely to occur in nature for ferromagnetic $J_{1}$, because smaller
anisotropies are required \cite{meta}. Interestingly, for ferromagnetic
small $J_{1}$ a very small gap with an astronomically large associated
correlation length has been found by field theory calculations and seems
consistent with density matrix renormalization group (DMRG) calculations 
\cite{itoi}.

The phase diagram of the model for $\Delta _{1}=\Delta _{2}>0$ has been
accurately determined by Nomura and Okamoto using the method of level
crossings of the excitation spectra, supported by results of conformal
field-theory and renormalization group \cite{nom}. This method has been also
used successfully to determine phase diagrams of electronic models \cite{nak}%
, and is related in some cases to jumps in topological numbers determined by
Berry phases \cite{ali,tor}. Using the method of level crossings, the phase
diagram has been extended recently to the region $\Delta _{1}=\Delta _{2}<0$ 
\cite{hir2}. However, to assume that $\Delta _{2}<0$ implies a different
sign of the $z$ component of $J_{2}$ with respect to those of the $x,y$
plane. This is an unrealistic anisotropy (this inconvenient is not present
for $J_{1}$ because the sign of its $x,y$ components can be changed rotating
every second spin in the chain by $\pi $ around the $z$ axis). In addition,
the above mentioned systems containing CuO chains with edge-sharing
plaquettes and Cu-O-Cu angle near 90${{}^{\circ }}$, are expected to lie
near the isotropic case $\Delta _{1}=-1,$ $\Delta _{2}=1$.

In this work we calculate the phase diagram of the system for $\Delta
_{2}=\left| \Delta _{1}\right| $ and $-1.2<\Delta _{1}<1.2$ using the method
of level crossings. This includes the interpolations between the {\it XX}
chain and the isotropic cases with ferro- or antiferromagnetic $J_{1}$.

\section{Model and method of level crossings}

The model is:

\begin{eqnarray}
H &=&\sum_{i}[J_{1}(S_{i}^{x}S_{i+1}^{x}+S_{i}^{y}S_{i+1}^{y}+\Delta
_{1}S_{i}^{z}S_{i+1}^{z})  \nonumber \\
&&+J_{2}(S_{i}^{x}S_{i+2}^{x}+S_{i}^{y}S_{i+2}^{y}+\Delta
_{2}S_{i}^{z}S_{i+2}^{z})],  \label{h1}
\end{eqnarray}
where $S_{i}^{\beta }$ is the $\beta $ component of the spin-1/2 operator at
site $i$.

The basic idea of the method of level crossings is that the properties of a
critical phase at low energies and large distances are ultimately determined
by a scale invariant fixed point \cite{nom}. Then, using conformal field
theory one can relate the excitation energy which corresponds to some
operator $A_{i}$ at site $i$ (for example a spin flip $S_{i}^{+}$, $%
S_{i}^{-} $), to the dependence of the correlation functions of this
operator with distance: 
\begin{equation}
E_{A}(L)-E_{g}(L)=\frac{2\pi vx_{A}}{L},\text{ }\left\langle
A_{i+d}A_{i}\right\rangle \backsim \frac{1}{d^{2x_{A}}}.  \label{2}
\end{equation}
Here $L$ is the length of the system, $v$ the spin-wave velocity, $E_{g}(L)$
the ground state energy, and $x_{A}$ the critical dimension for the
excitation $A$. For example if $J_{2}=0$, the {\it XXZ} model was exactly
solved and its properties are well known \cite{joh,gog}. The long distance
dependence of the main correlation functions is

\begin{equation}
\left\langle S_{i+d}^{+}S_{i}^{-}\right\rangle \backsim (-1)^{d}r^{-1/K},%
\text{ }\left\langle S_{i+d}^{z}S_{i}^{z}\right\rangle \backsim
(-1)^{d}r^{-K}  \label{3}
\end{equation}
with 
\begin{equation}
K=\frac{\pi }{\pi -\arccos (\Delta _{1})}  \label{4}
\end{equation}
if $\left| \Delta _{1}\right| \leq 1$. For $\left| \Delta _{1}\right| <1,$
the minimum excitation corresponds to a spin flip with wave vector $\pi .$
The ground state has total spin projection $S^{z}=\pm 1,$ its wave vector
differs from that of the ground state in $\pi $, and its parity under
inversion $P$ is opposite to that of the ground state \cite{nom}. The matrix
element of $S_{i}^{+(-)}$ between the ground state and the first excited
doublet is different from zero. Then, the correlation functions $%
\left\langle S_{i+d}^{+}S_{i}^{-}\right\rangle $ are the dominant ones at
large distances.

At $\Delta _{1}=1$, the above mentioned doublet crosses with the excited
state with $S^{z}=0$, (with wave vector differing from the ground state in $%
\pi $, and $P$ and parity under spin reversal $T$ opposite to that of the
ground state). Then, for $\Delta _{1}>1$, $\left\langle
S_{i+d}^{z}S_{i}^{z}\right\rangle $ are the dominant correlations at large
distances. At $\Delta _{1}=1$, a gap opens and conformal invariance ceases
to be valid, but for a sufficiently small gap so that correlation length is
much larger than the size of the system, Eqs. (\ref{2}) are still expected
to hold.

In summary inside a critical phase or near its boundaries, the quantum
numbers of the first excited state determine the dominant correlations at
large distances, and therefore the crossings of the excited states determine
the boundaries.

For small $J_{2}/J_{1}$, the use of renormalization group allows to
determine the size dependence of the crossings and to extrapolate them very
accurately to the thermodynamic limit \cite{nom}.

\section{Results}

The phase diagram we obtain in the plane $\alpha =J_{2}/J_{1}$ as a function
of $\Delta _{1},$ with $\Delta _{2}=\left| \Delta _{1}\right| $ is
represented in Fig. 1. The transition between dimer and spin fluid I phases
for $\Delta _{1}\geq 0$ has been obtained before by Nomura and Okamoto \cite
{nom}. The spin fluid I phase is the well known ground state of the model
for $\alpha =0,$ $\left| \Delta _{1}\right| <1$. The dimer or spin Peierls
phase has a spin gap and broken translational symmetry (the unit cell is
doubled) in the thermodynamic limit. The first excited state in finite
systems has the same quantum numbers $P$ and $T$ as the ground state, but
its wave vector differs by $\pi $ \cite{nom}. The new results for the spin
fluid I{\bf -}dimer transition are displayed in Fig. 2. The transition
points (open squares) were accurately determined extrapolating the crossing
of the first excited state using data for chain lengths $L=12,$ $16$ and $20$%
. As explained earlier \cite{nom} the crossing point has a $1/L^{2}$ size
dependence. We have verified this also for negative $\Delta _{1}$ (see Fig.
3). However, in the region denoted by {\bf ? }in Fig. 3 (constructed with
data for $L=16$), the first excited state has opposite $P$ and $T$ as the
ground state. This is unexpected for the dimer phase. This might be a
finite-size effect related with the high degeneracy at the point $\Delta
_{1}=-\Delta _{2}=-1$, $\alpha =0.25$ \cite{meta,rub}. The other possibility
is that a novel phase exists there, but this seems unlikely.

After the transition to the dimer phase, if $\alpha $ is increased further,
the crossing of excitation spectra indicates another transition to a phase
in which the first excited state has $S^{z}=\pm 1$ and wave vector $%
k_{c}=\pi /2$. We denote this phase as ``spin fluid II '' (between quotation
marks), because, as we discuss below, its nature is not established for all
values of $\Delta _{i}$. For $\Delta _{i}=0$, from field theory results one
might expect a spin fluid phase for large $\alpha $ \cite{ner}. One also
expects that $k_{c}$ is actually incommensurate and slightly different from $%
\pi /2$ \cite{bur,whi,ner,inco}. However, we are not able to detect this
difference using periodic boundary conditions. While the use of twisted
boundary conditions has given interesting results concerning this
incommensurability \cite{inco}, to implement them in the level crossing
method is time consuming and present technical complications, which we avoid
in this work.

As shown in Fig. 4, the level crossing for the dimer-spin fluid II
transition does not follow a $1/L^{2}$ dependence, and the extrapolation to
the thermodynamic limit is not so accurate as for the dimer-spin fluid I
transition. We have used a quadratic extrapolation in $1/L^{2}$, using the
data for $L=8,$ $12,$ $16$ and $20$. Field-theoretical results suggest that
there is a small gap and therefore no spin fluid phase for $\Delta
_{1}=\Delta _{2}=1$, and $J_{2}>>J_{1}$ \cite{whi,all,cab}. For $\Delta
_{1}=-1$, $\Delta _{2}=1$, a tiny gap is predicted with an associated
astronomically large length scale \cite{itoi}. A direct extrapolation of the
gap from our finite-size results has an error larger or of the order of the
gap itself and therefore, it is not able to establish if the gap is open or
not. However, in principle we can check if a given phase is a spin fluid
phase by other methods. For example, the energy per site should vary with
system size as: 
\begin{equation}
e(L)=e(\infty )-\frac{\pi }{6L^{2}}v_{s}  \label{ene}
\end{equation}
where $v_{s}$ is the spin velocity (or the sum of spin velocities if there
were several types of low-energy excitations). We have checked that Eq. (\ref
{ene}) is not satisfied inside the dimer phase. A fit of the energies for $%
L=12,$ $16,$ $20$ and $24$ with Eq. (\ref{ene}) give an error of $v_{s}$
which is of the order of $v_{s}$ itself. Instead, inside both ``spin fluid''
phases the fit is good. The error in $v_{s}$ is of order of 0.1\% inside
spin fluid I, and of order of 1\% inside spin fluid II. Within spin fluid I,
we also find that several relations derived from conformal field theory
hold, and are consistent within them within $\backsim 5\%$. The spin
velocity can be calculated as: 
\begin{equation}
v_{1}=\frac{E(2\pi /L)-E(0)}{2\pi /L},  \label{v}
\end{equation}
where $E(q)$ is the lowest energy in the $S^{z}=0$ sector. The extrapolation
of $v_{1}$ to the thermodynamic limit using a quadratic polynomial in $1/L$
is only slightly less than $v_{s}$ obtained fitting Eq. (\ref{ene}). Also
from the numerical calculation of the susceptibility $\chi $ and spin
stiffness $D_{s}$: 
\begin{equation}
\chi _{s}=\frac{1}{\frac{\partial ^{2}e}{\partial m^{2}}}=\frac{1}{%
L[E(1)+E(-1)-2E(0)]}  \label{xi}
\end{equation}
\begin{equation}
D_{s}=\frac{L}{2}\frac{\partial ^{2}E}{\partial \phi ^{2}}  \label{fi}
\end{equation}
where here $E(S^{z})$ is the lowest energy for total spin $S^{z}$, and $\phi 
$ is a flux opposite for spin up and down, \cite{inco} we can obtain $K$ in
two independent ways from the relations \cite{gog}: 
\begin{equation}
\chi _{1}=\frac{K}{2\pi v_{1}},\text{ }D_{1}=\frac{Kv_{1}}{4\pi }  \label{k}
\end{equation}
The correlation exponents obtained from these two equations and Eqs. (\ref{v}%
), (\ref{xi}), (\ref{fi}) with $\chi _{1}=\chi _{s},$ $D_{1}=D_{s}$ also
agree within 2\% to 10\% in the spin fluid phase I. The agreement improves
with increasing $\Delta _{1}$.

A simple analysis based on Eqs. (\ref{xi}), (\ref{fi}), (\ref{k}) is not
possible in the ``spin fluid II'' phase. Eqs. (\ref{ene}) and (\ref{v})
suggest that the low energy properties of this phase are given by {\em two}%
{\bf \ }free bosonic theories (this is of course true if $J_{1}=0$): $%
v_{s}=v_{1}+v_{2}$, with $v_{2}$ slightly larger than $v_{1}$ (see Fig.
5(a)). This is in principle reasonable, since if there are two branches of
low-energy excitations, Eq. (\ref{v}) should give the smallest velocity.
Also one expects that Eq. (\ref{fi}) gives the sum of both spin stiffness: 
\begin{equation}
D_{s}=\frac{K_{1}v_{1}+K_{2}v_{2}}{4\pi }  \label{d}
\end{equation}
The case of the susceptibility is more delicate. From thermodynamics, the
total susceptibility should be the sum of those of both branches. However,
our numerical study suggests that when one flips only the spin, it goes to
the branch of lowest velocity.

The results of field theory in the case $J_{2}\gg J_{1}$ \cite{all,ner} (and
also those of the Hubbard ladder for small interchain hopping \cite{hub,note}%
) suggest that the effective low-energy theory splits in two sectors,
symmetric $\varphi _{+}$ and antisymmetric $\varphi _{-}$, and in general,
at least the latter is gapped. For $\Delta _{1}=\Delta _{2}=0$, $\varphi
_{+} $ is massless and $\varphi _{-}$ has a gap which depends on $J_{2}/J_{1}
$ as a power law \cite{ner,inco}. In the isotropic case $\Delta _{1}=\Delta
_{2}=1 $, both sectors have an exponentially small gap \cite{all}. These
field theory results are inconsistent with the hypothesis of two branches of
massless excitations in the ``spin fluid II'' phase. To check this
hypothesis we investigated the consistency of Eqs. (\ref{ene}) with $%
v_{s}=v_{1}+v_{2}$, (\ref{v}), (\ref{d}) and an expression for the total
susceptibility, for some values of $K_{1}$ and $K_{2}$ in the above
mentioned cases ($\Delta _{1}=\Delta _{2}=0$, and $\Delta _{1}=\Delta _{2}=1$%
). We do not find reasonable results. For example, eliminating $K_{i}$ from
these equations, sometimes one negative result arises. The most plausible
explanation is that at least one of the modes is gapped, but the system is
not large enough to detect it in the size dependence of the energy (Eq. (\ref
{ene}) is only valid for sufficiently large $L$). In fact, some of our
fittings suggest that the magnitude of the slope of $e$ vs $1/L^{2}$ is
decreasing with system size.

To end this section, we discuss the observed phases for $1.2>\left| \Delta
_{i}\right| >1$. For $\Delta _{1}<-1,$ the ground state changes from a fully
polarized ferromagnet ($S^{z}=L/2$) to $S^{z}=0$ at the solid squares of
Fig. 1. Actually, there is a small region of intermediate $S^{z}$ between
both phases, but it decreases with system size an seems to disappear in the
thermodynamic limit. In the phase with $S^{z}=0,$ the first excited state
has wave vector $\pi /2$. This fact, a study of correlation functions \cite
{som} and a trivial analysis of energies in the classical limit $-\Delta
_{1}=\Delta _{2}\rightarrow \infty $, suggests that the phase has long-range
AF order of the type $\uparrow \uparrow \downarrow \downarrow \uparrow
\uparrow ...$ (denoted as AFII). The same features denoting the presence of
the AFII phase, are also found for $\Delta _{i}>1$ and large $\alpha $. We
expect that for $\left| \Delta _{i}\right| >1$, a spin wave approximation
can describe qualitatively the essential physics.

\section{Discussion}

Using the method of level crossings, we have calculated the phase diagram of
the {\it XXZ} model with next-NN exchange (Eq. (\ref{h1})) for $\Delta
_{2}=\left| \Delta _{1}\right| $. For the dimer-spin fluid I transition our
results are quite robust and extend previous results \cite{nom,hir2} to the
case $1\leq \Delta _{1}=-\Delta _{2}<0$. In this region, the transition
takes place near $\alpha \backsim 0.3$. For larger values of $\alpha $ the
method predicts a transition from the dimer phase to a second spin fluid
phase. The transition is at $\alpha _{c}=1.42$ for $\Delta _{1}=\Delta
_{2}=0 $, in qualitative agreement with $\alpha _{c}\backsim 1.26$ obtained
in Ref. \cite{hik}. For these values of $\Delta _{i}$, field theoretical
results predict the existence of this phase \cite{ner}.

For $\Delta _{1}=-1$, a tiny gap with an astronomically large length scale
is predicted in Ref. \cite{itoi}. Such a length is of course much larger
than the system sizes we use, and our finite size results are consistent
with those of Ref. \cite{itoi}.

For $\Delta _{1}=\Delta _{2}=1$, we have tried to interpret the
thermodynamic properties of the ``spin fluid II'' phase as a sum of two
independent fluids, but the results are not consistent. We conclude then
that at this point, the extrapolation of the level crossings is not reliable
and in agreement with other calculations \cite{whi,all,cab}, the gap
persists for large $J_{2}$.

From the parameters estimated for several compounds near the isotropic limit 
$\Delta _{1}=-\Delta _{2}=-1:$ La$_{6}$Ca$_{8}$Cu$_{24}$O$_{41}$ ($\alpha
\backsim 0.36$), Li$_{2}$CuO$_{2}$ ($\alpha \backsim .62$) and Ca$_{2}$Y$%
_{2} $Cu$_{5}$O$_{10}$ ($\alpha \backsim 2.2$) \cite{miz} our results
indicate that these compounds lie outside the usual spin fluid I phase. They
are in the AF2 phase if $\Delta _{1}<-1$. If $\Delta _{1}\geq -1$, La$_{6}$Ca%
$_{8}$Cu$_{24}$O$_{41}$ is clearly in the dimer phase. One expects a very
small, perhaps unobservable gap for Ca$_{2}$Y$_{2}$Cu$_{5}$O$_{10}$ and
spin-spin correlations with wave vector near $\pi /2$. Li$_{2}$CuO$_{2}$
lies in between. The interchain interactions might affect this scenario.

\section*{Figure captions}

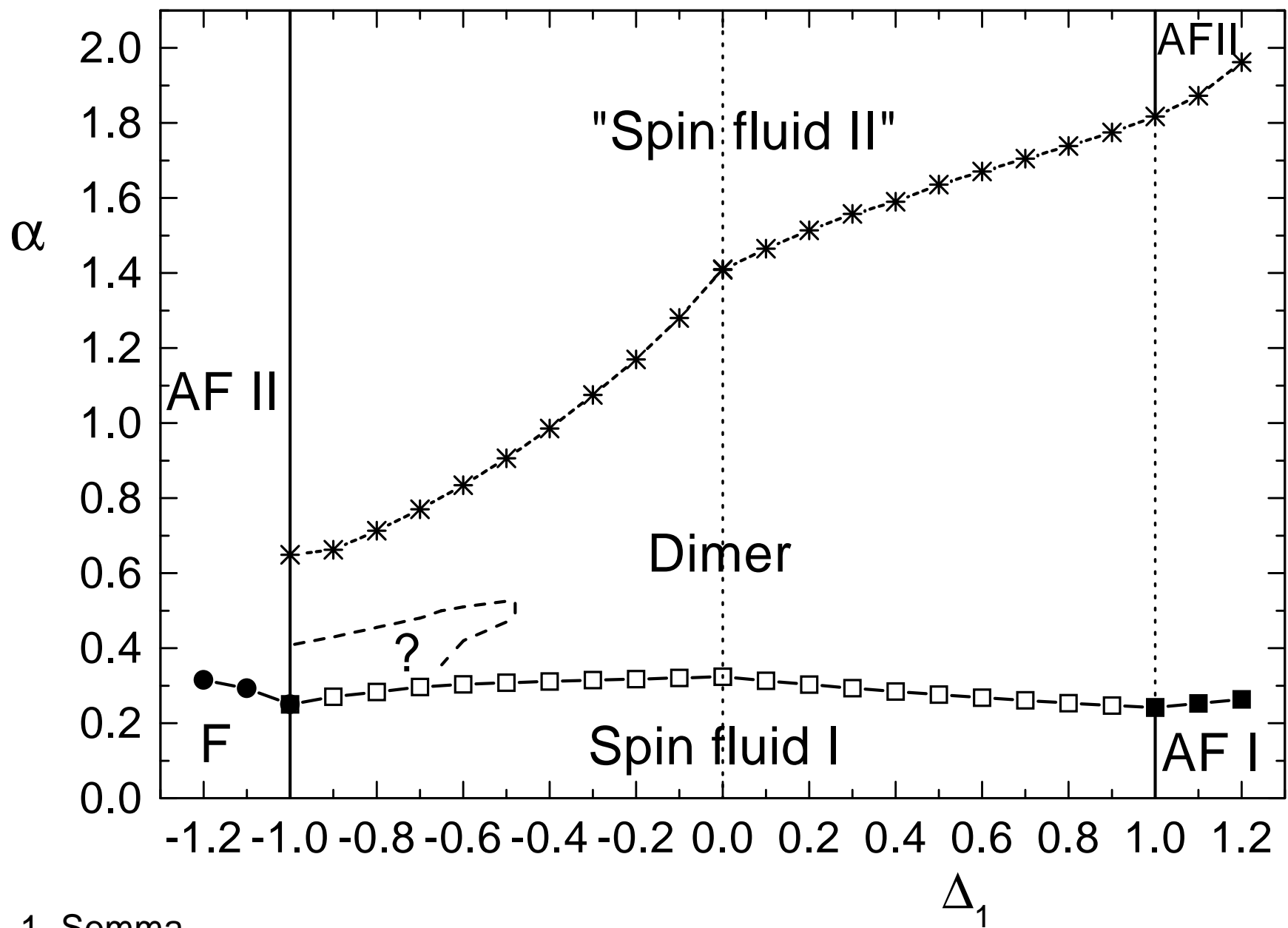
{\bf Fig.1. } Phase diagram of the model Eq. (\ref{h1}) in the $\Delta _{1},$
$\alpha =J_{2}/J_{1}$ plane, keeping $\Delta _{2}=\left| \Delta _{1}\right|
. $ F denotes the fully magnetized ferromagnetic phase, and AF2 a phase with
long range order $\uparrow \uparrow \downarrow \downarrow ...$ .

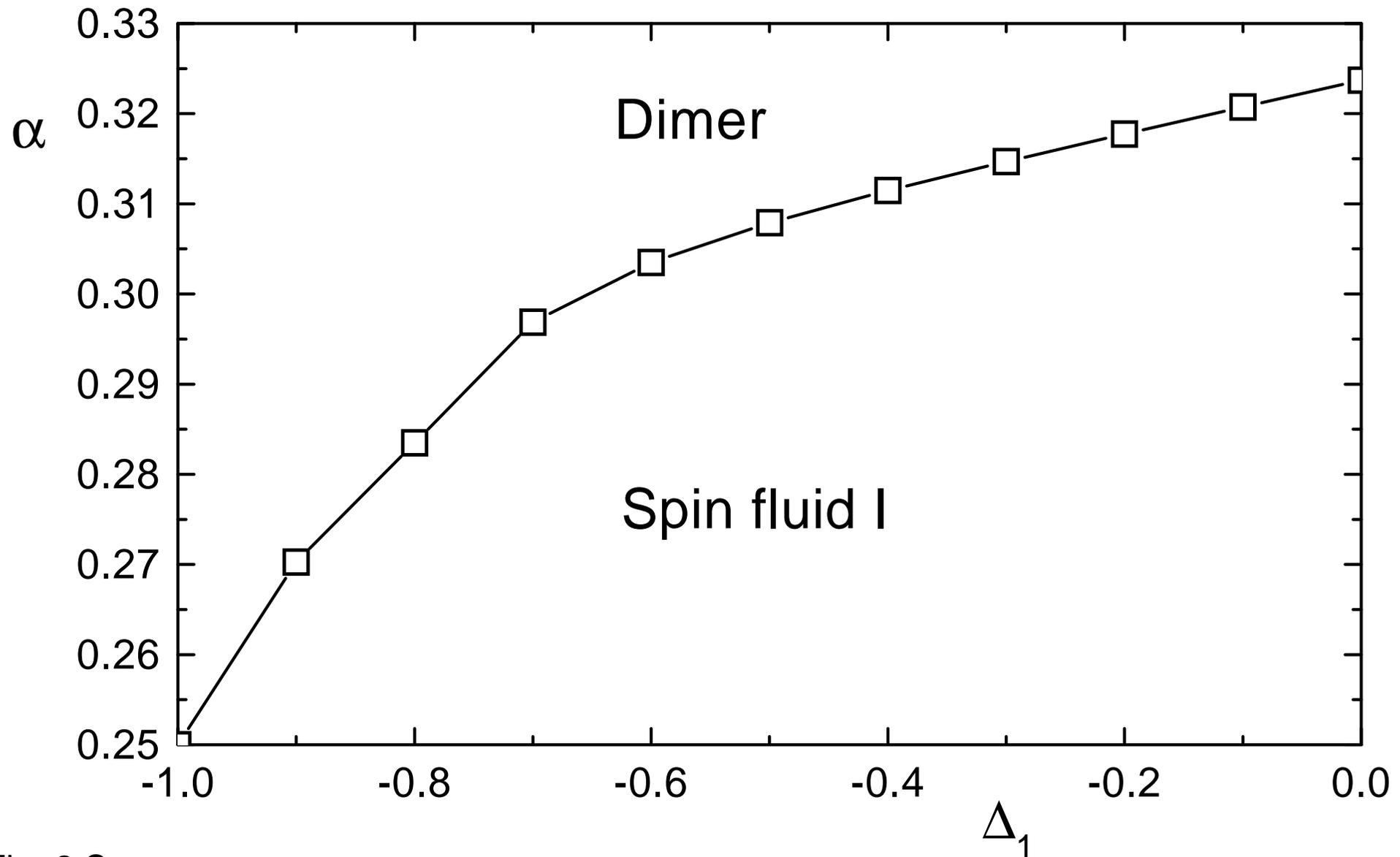
{\bf Fig.2. } Details of the boundary of the spin fluid I phase of Fig.1 for
negative $\Delta _{1}$.

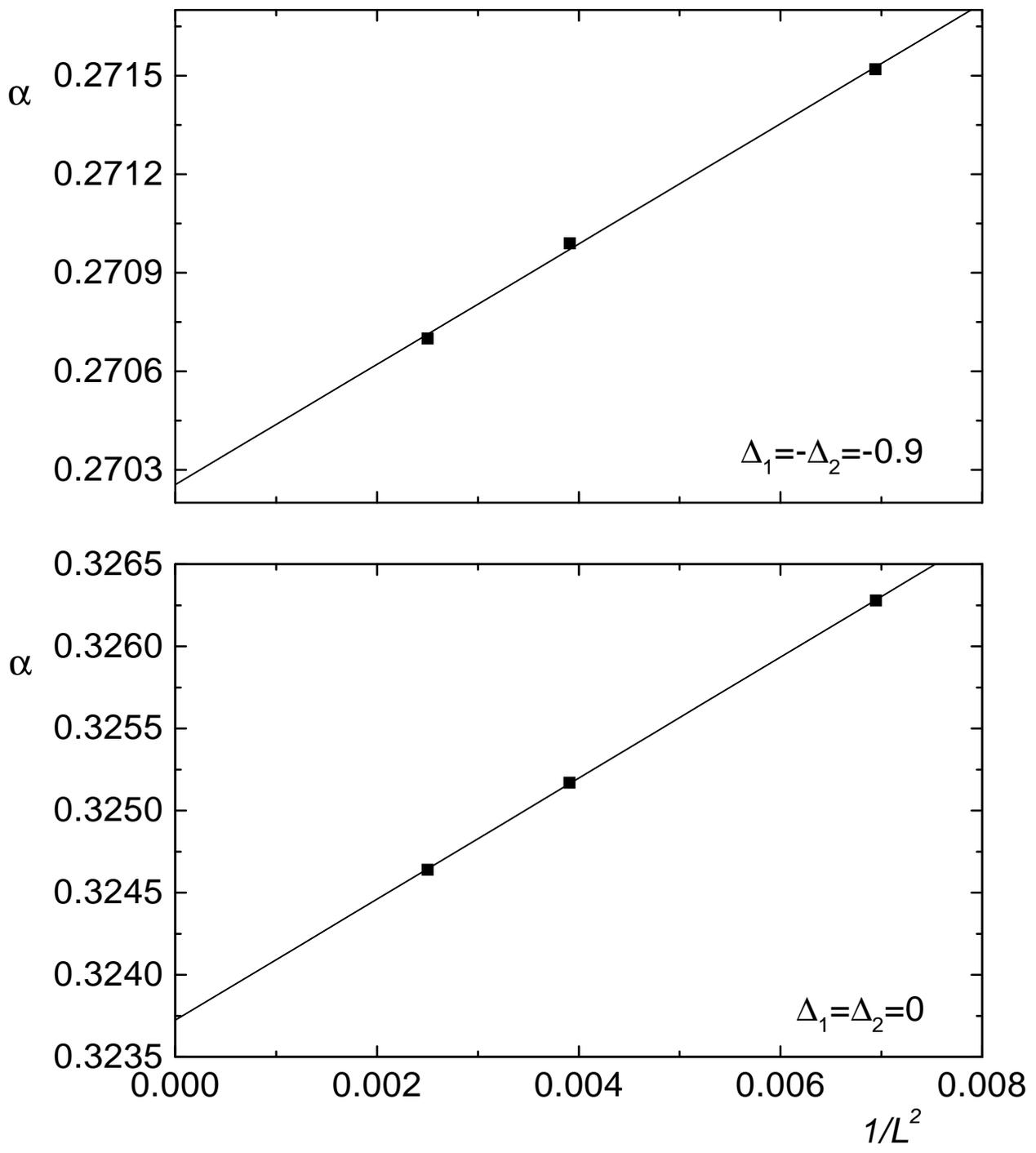
{\bf Fig.3. } Size dependence of the crossing between dimer and spin fluid I
excitations.

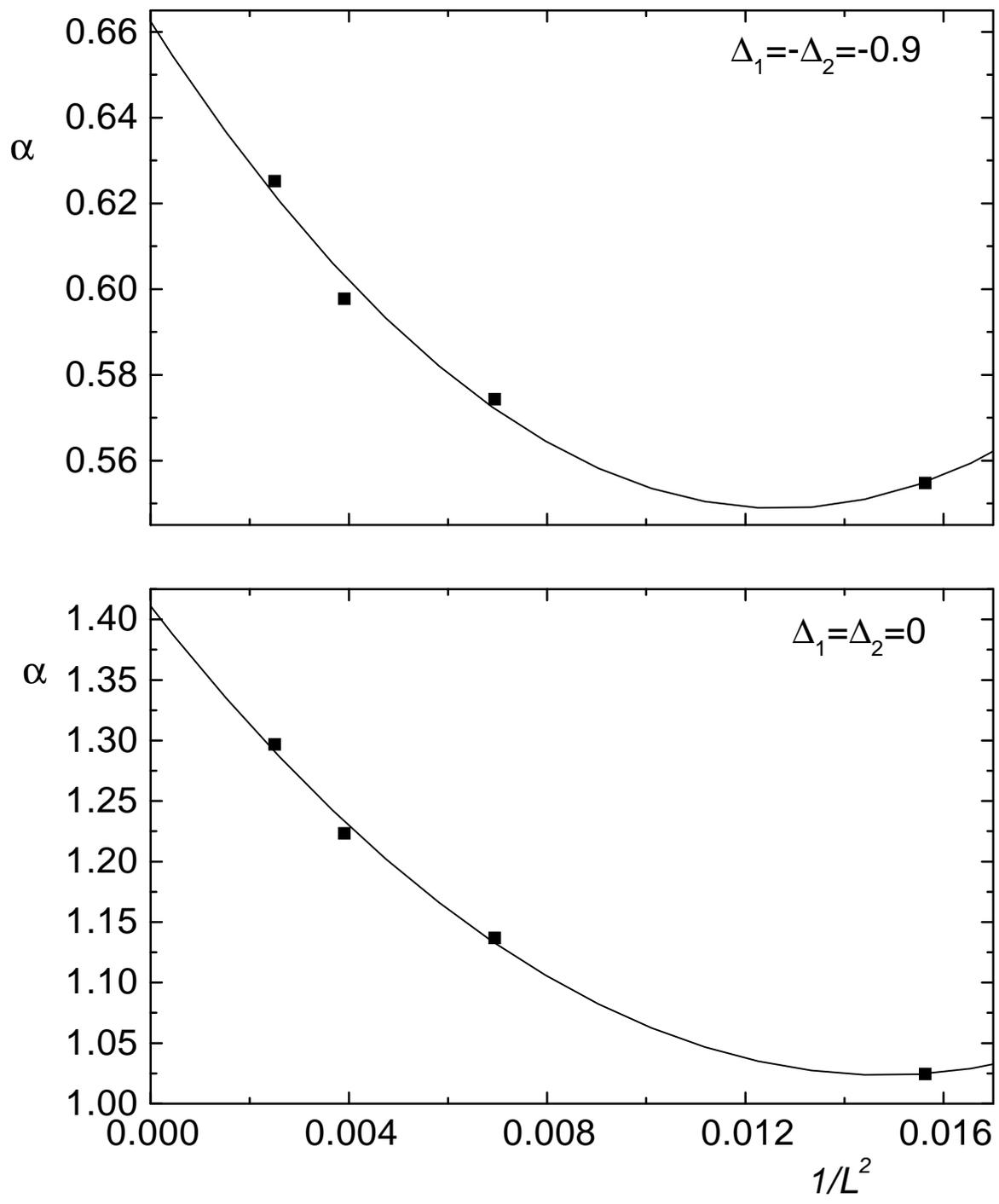
{\bf Fig.4. } Size dependence of the crossing between dimer and ``spin fluid
II'' excitations.


\begin{references}
\bibitem{has}  M. Hase, I. Terasaki, and K. Uchinokura, Phys. Rev. Lett. 
{\bf 70}, 3651 (1993).

\bibitem{mot}  N. Motoyama, H. Eisaki and S. Uchida, Phys. Rev. Lett. {\bf 76%
}, 3212 (1996).

\bibitem{col}  R. Coldea, D.A. Tennant, R.A. Cowley, D.F. McMorrow, B.
Dorner and Z. Tylczynski, Phys. Rev. Lett. {\bf 79}, 151 (1997).

\bibitem{miz}  Y. Mizuno, T. Tohyama, S. Maekawa, T. Osafune, N. Motoyama,
H. Eisaki, and S. Uchida, Phys. Rev. B {\bf 57}, 5326 (1998).

\bibitem{bur}  R. Bursill, G.A. Gehring, D.J.J. Farnell, J.B. Parkinson, T.
Xiang and C. Zeng, J. Phys. Cond. Mat {\bf 7}, 8605 (1995).

\bibitem{whi}  S. R. White and I. Affleck, Phys. Rev. B {\bf 54}, 9862
(1996).

\bibitem{all}  D. Allen and D. S\'{e}n\'{e}chal, Phys. Rev. B {\bf 55}, 299
(1997).

\bibitem{ner}  A. A. Nersesyan, A. O. Gogolin, and F. H. L. E\ss ler, Phys.
Rev. Lett {\bf 81}, 910 (1998).

\bibitem{inco}  A.A. Aligia, C.D. Batista and F.H.L. E\ss ler, Phys. Rev. B 
{\bf 62}, 3259 (2000).

\bibitem{hik}  T. Hikihara, M. Kaburagi and H. Kawamura, cond-mat/0007095.

\bibitem{ger}  C. Gerhardt, K.-H. M\={u}tter and H. Kr\"{o}ger, Phys. Rev. B 
{\bf 57}, 11 504 (1998).

\bibitem{hir1}  S. Hirata, cond-mat/9912066.

\bibitem{meta}  A. A. Aligia, Phys. Rev. B {\bf 62}, 14 402 (2001).

\bibitem{cab}  D. Cabra, A. Honecker, and P. Pujol, Eur. Phys. Jour. B {\bf %
13}, 55 (2000).

\bibitem{hir2}  S. Hirata and K. Nomura, Phys. Rev. B {\bf 61}, 9453 (2000).

\bibitem{itoi}  C. Itoi and S. Qin, cond-mat/0006155.

\bibitem{nom}  K. Nomura and K. Okamoto, J. Phys. A {\bf 27}, 5773 (1994).

\bibitem{nak}  M. Nakamura, Phys. Rev. B {\bf 61}, 16 377 (2000).

\bibitem{ali}  A. A. Aligia, K. Hallberg, C. D. Batista, and G. Ortiz, Phys.
Rev. B {\bf 61}, 7883 (2000).

\bibitem{tor}  M. E. Torio, A. A. Aligia, K. Hallberg, and H. A. Ceccatto,
Phys. Rev. B {\bf 62}, 6991 (2000).{\bf \ }

\bibitem{joh}  J. D. Johnson, S. Krinsky, and B. M. Mc Coy, Phys. Rev. A 
{\bf 8}, 2526 (1973).

\bibitem{gog}  A. Gogolin, A. A. Nersesyan and A. M. Tsvelik, {\it %
Bosonization and strongly correlated systems }(Cambridge, New York, 1998).

\bibitem{rub}  T. Hamada, J. Kane, S. Nakagawa, and Y. Natsume, J. Phys.
Soc. Jpn {\bf 57}, 1891 (1988).

\bibitem{hub}  L. Balents and M.P.A. Fisher, Phys. Rev. B {\bf 53}, 12 133
(1996); M. Fabrizio, Phys. Rev. B {\bf 54}, 10 054 (1996).

\bibitem{note}  The Hubbard zig-zag ladder at half filling in the limit of
large on-site Coulomb repulsion reduces to the isotropic case $\Delta
_{1}=\Delta _{2}$ of the present model.

\bibitem{som}  R. D. Somma and A. A. Aligia, Solid State Commun. {\bf 117},
273 (2001).
\end{references}
\end{document}